\documentclass{aastex}
\usepackage{emulateapj5}

\newcommand{\kms}    {km$\:$s$^{-1}$}
\newcommand{\Ms}     {$M_\odot$}

\def\feiiq{\rm Fe{\sc ii }$\lambda$4570\/}
\def\feii{{Fe\sc{ii}}\/}
\def\hb{{\sc{H}}$\beta$\/}

\def\hbbc{{\sc{H}}$\beta_{\rm BC}$\/}

\def\rfe{R$_{\rm FeII}$}

\slugcomment{The Astrophysical Journal, submitted  2002 February 26, accepted 2002 April 24}

\shorttitle{From Quasars to Nano-Quasars} \shortauthors{Zamanov \&
Marziani}

\begin{document}


\title{Searching for the Physical Drivers of Eigenvector~1: \\
 From Quasars to Nano-Quasars}


\author{R. Zamanov and P. Marziani }
\affil{Osservatorio Astronomico di Padova, INAF, Vicolo
dell'Osservatorio 5,
             I-35122 Padova, Italy }
\email{zamanov@pd.astro.it ,  marziani@pd.astro.it}



\begin{abstract}
We point out an analogy between two accreting white dwarfs with jets
(CH~Cyg  and MWC~560) and  powerful quasars. In spite of the
enormous  difference in the mass of the central object (a factor
$\sim$10$^{7}$), the emission lines are strikingly similar to those
of I~Zw~1 (the prototype ``Narrow Line Seyfert 1'' nucleus whose
spectrum is widely used as an FeII template for almost all quasars).
The spectral similarity give us the unique possibility to consider
the optical Eigenvector~1 diagram using objects less massive by a
factor of millions. Our results reinforce the interpretation of the
``Eigenvector 1 correlations'' found for low redshift quasars as
driven mainly by the source luminosity  to central compact object
mass ratio(L/M). The accreting white dwarfs CH~Cyg and MWC~560,
their jets and emission lines, may well represent the low energy,
non~relativistic end of the accretion phenomena, which encompass the
most powerful quasars and the microquasars. The remarkable
similarities suggest that they may be legitimately considered
``nano-quasars".
\end{abstract}




\section{Introduction}

The accretion processes have a lot of similarities in
spite of the differences in the type and the mass of the accreting
object. In the last years a few microquasar sources have been
discovered. They are galactic X-ray binaries in which a black hole
(BH) or neutron star is accreting from the companion star,
producing jets and even superluminal motion (Mirabel \& Rodriguez,
1999). The  accreting stellar mass compact objects give us a tool
to investigate phenomena in active galactic nuclei (AGN)
at much lower energy and much shorter time scale.

The aim of this letter is to show striking similarities between the
emission lines of two accreting white dwarfs and the  emission lines
coming from the broad line region of the active galactic nuclei
(AGN), in spite of the mass difference (a typical BH in AGN
has a mass $\sim$10$^6$--10$^9$\Ms  \ and the white dwarf
in the interacting binaries  $\sim$1\Ms). A comparison
between such different objects can give us a better
understanding of the accretion/ejection phenomenology, as well as of
the so called ``Eigenvector~1 correlations" that seem to be
fundamental for AGN interpretation. We think it is appropriate to
call these two accreting white dwarfs of this paper ``nano-quasars"
because they seem to represent the very low energy analog of quasars
and microquasars.

\section{Emission lines similarities}

\begin{figure*}[htb]
 \mbox{}
 \vspace{19.0cm}
 \includegraphics{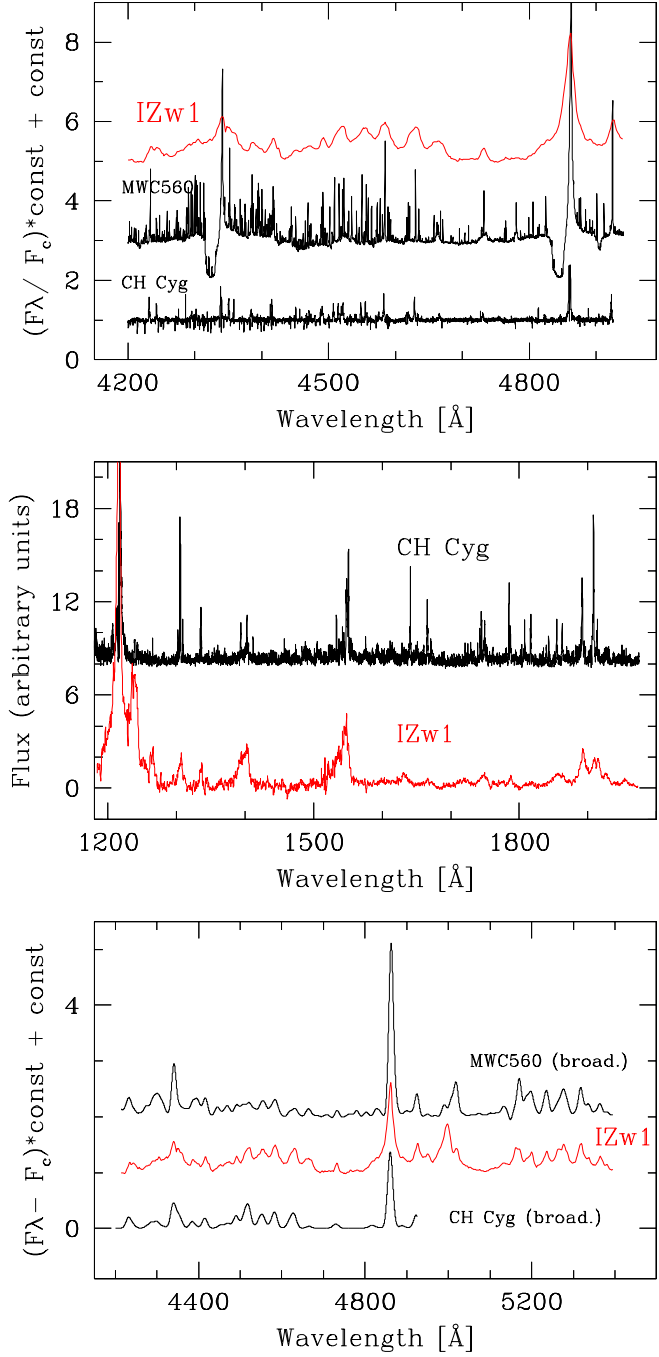}
  \caption[]{ A comparison between the spectra of the interacting binaries
  CH~Cyg, MWC~560 and the low redshift quasar I~Zw~1.
  Upper Panel:   optical  spectra;  Middle Panel:  the UV region; Lower Panel:
  the optical spectra of MWC~560 and CH~Cyg after broadening and scaling.
  A clear similarity
  between  the emission lines is visible in all panels. 
 }
 \label{SPEC}
\end{figure*}

CH~Cyg and MWC~560 are interacting binary stars in which a white
dwarf accretes matter from the wind of a red giant. Their spectra
and the spectrum of I~Zw~1 are plotted in Fig.\ref{SPEC}. I~Zw~1 is
a ``Narrow-Line" Seyfert 1 (NLSy1) nucleus, well known because of his
strong \feii\ emission and relatively narrow lines
(FWHM$\sim$1000~\kms), widely used as template for subtraction of
the FeII complex in the H$\gamma$ - \hb\   region of quasar spectra.
In the upper panel of Fig. \ref{SPEC} we show the spectral region
$\lambda$4200--$\lambda$4900 \AA, and in the middle panel the UV
region. The optical spectrum  of I~Zw~1 was obtained at the 1.52m
ESO telescope at La Silla, and the UV was retrieved from the Hubble
(HST) archive. The  optical spectrum of MWC~560 is the average of
the FEROS public archive spectra observed in the period November -
December 1998 at ESO (Kaufer et 1999, Schmid 2001). Both optical and
UV spectra of CH~Cyg were obtained in 1984,
when the star underwent jet activity. The optical spectrum of CH Cyg
is from the plate archive of the Bulgarian National Astronomical
Observatory ``Rozhen" (observed on 10 July 1984). The UV spectrum
have been retrieved from the International Ultraviolet Explorer database
(observed on 23 Jan 1985).

A clear similarity between the emission lines can be seen in Fig.1.
Practically almost every emission line visible in the  spectrum of
I~Zw~1 has corresponding  features in the spectra of CH~Cyg and
MWC~560. An obvious similarity is visible also between the UV
spectrum of CH~Cyg and the one of I Zw 1. We note in passing that
similar emission lines are also visible in the spectrum of XX Oph
(Kolev \& Tomov 1993), where the accreting object is probably a main
sequence star (Evans et al. 1993). Despite the general similarity
some differences are visible as well. In MWC 560 these are the
absorption components in the Balmer lines. These absorptions are due
to the jet coinciding with the line of sight (Tomov et al. 1990).
They are also dominating the UV spectrum of MWC~560 (not shown here)
making it quite different from the ones of CH~Cyg and I~Zw~1. In the
optical spectrum of CH~Cyg numerous absorptions due to the
photosphere around the white dwarf are visible.

In Fig.\ref{SPEC} (lower panel) the optical emission line spectra of
MWC~560 and CH~Cyg are shown after having been continuum subtracted,
scaled and broadened. This standard procedure  is widely used for
the emission line measurements of AGN, using I~Zw~1 itself as a
template. After this processing, good identity is achieved with
the spectrum of I~Zw~1.
Our best fit corresponds to a width FWHM(FeII$_{opt}$)$=$970$\pm$90~\kms.

The hydrogen emission lines as well as the FeII emissions of AGN are
emitted from the so-called broad line region (BLR). This region is
thought to be within $\la$1~pc from the central BH. Its structure is
poorly understood as yet. The clear similarity between the emission
lines means that, in objects like MWC~560 and CH~Cyg,
we are observing a scaled down version of the  quasar BLR.

It worth noting that the interacting binaries, in which  a white
dwarf accretes material from the wind of a red giant (usually
classified as symbiotic stars) are strongly variable objects. For
CH~Cyg and MWC~560 we show that their spectra are similar to
low-redshift quasars in moments when jet activity has been detected
(see \S \ref{jets}).

\section{The Eigenvector-1  Diagram }

\begin{figure*}[htb]
  \includegraphics[width=9.0cm]{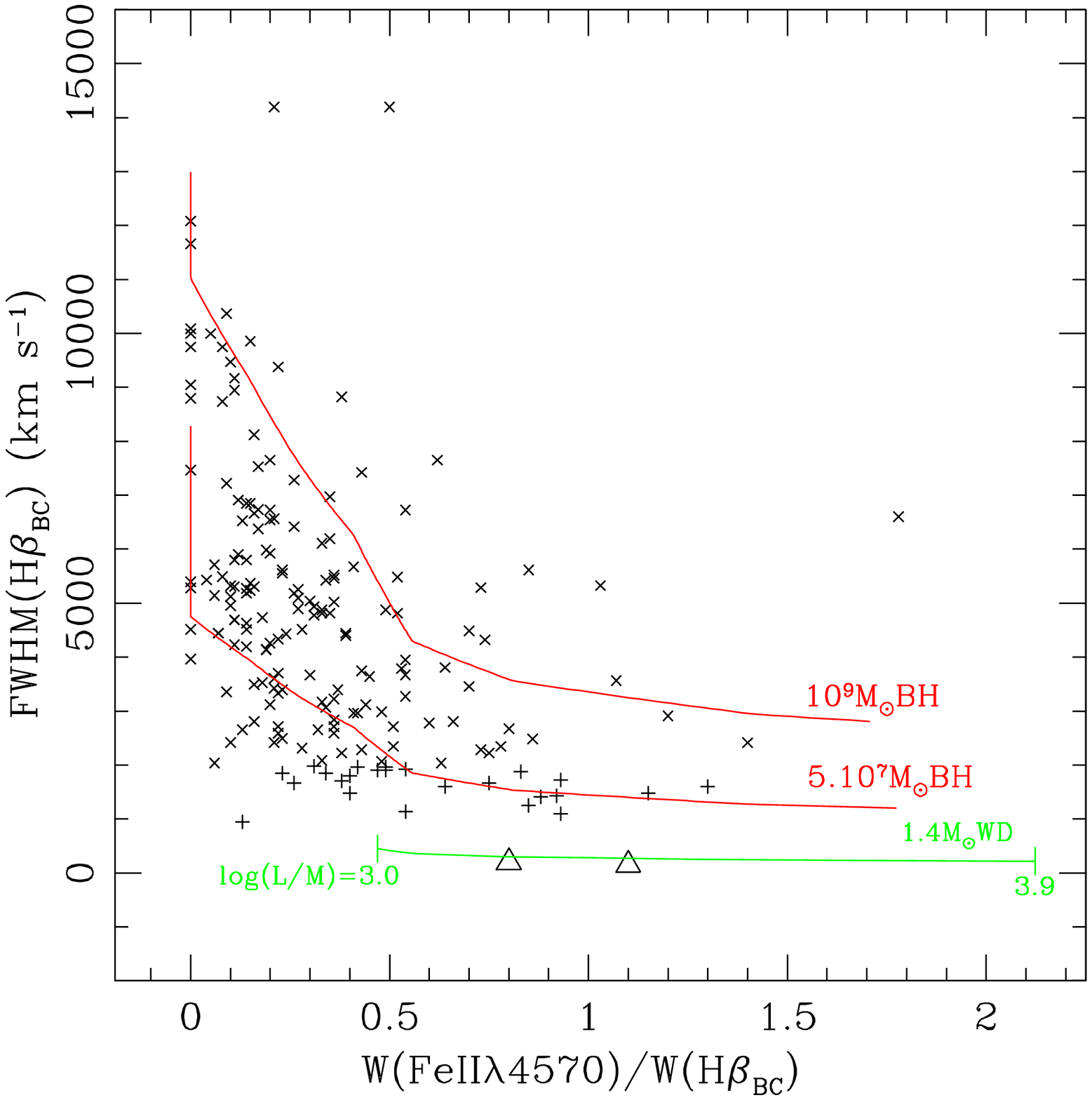}
  \includegraphics[width=9.0cm]{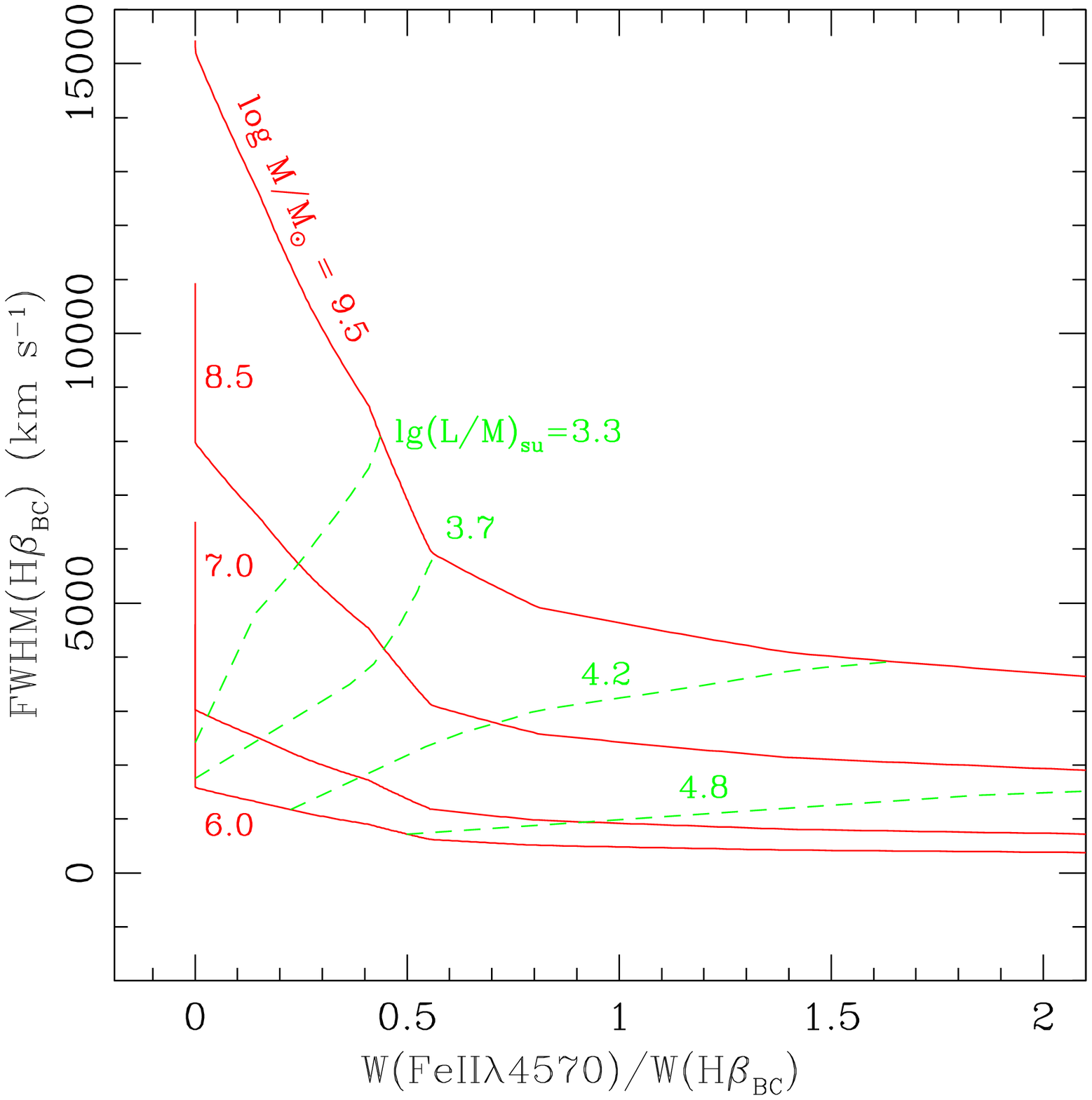}
  \caption{The FeII-H$\beta$ Eigenvector-1 diagram. Abscissa is
  rest frame equivalent width ratio between  \feiiq\ and \hbbc;
  ordinate is FWHM(\hbbc) in \kms.
  Left Panel: AGN from the sample of Sulentic et al. (2002)
  ($\times$ and $+$, $+$\ refers to NLSy1s),
  MWC~560 (located at \rfe=1.1) and CH~Cyg (\rfe=0.8)
  (triangles). The lines are plotted  (from top to bottom) for
  M$_{BH}=10^9$~\Ms, M$_{BH}=5\times10^7$~\Ms, and white dwarf
  mass M$_{WD}=1.4$. The line for 1.4\Ms WD is calculated in the limits
  $log(L/M)_{su}=$3.0 - 3.9.
  Right Panel: Theoretical grid for quasars in the optical E1
     diagram. The solid lines are plotted for fixed
     BH mass $log(M/M_\odot)=$9.5, 8.5, 7.0, 6.0. The dashed lines are for
     fixed L/M ratio $log(L/M)_{su}=$3.3, 3.7, 4.2, 4.8.
     \label{Fe-Hb} }
\end{figure*}


During the last decade, several investigations of AGN emission lines
emphasized the importance of a set of correlations conventionally
called ``Eigenvector-1" (hereafter E1) and related to principal
component analysis of the properties of PG quasars Boroson \& Green
(1992). In the most important diagram  associated to E1 (Sulentic
2000), the X-axis is the ratio between the equivalent width of the
FeII complex in the range 4434--4684 \AA \ and the equivalent width
of the H$\beta$\ broad component (\hbbc), namely R$_{FeII} =$
W(FeII)/W(H$\beta _{\rm BC}$). The Y-axis is the FWHM of \hbbc. We
measured for CH Cyg   W(FeII)=4.5\AA, W(H$\beta$)=5\AA, and
FWHM(H$\beta$)=200~\kms; and for  MWC~560  W(FeII)=34\AA,
W(H$\beta$)=31\AA, and FWHM(H$\beta$)=110~\kms.

The diagram is shown in Fig. 2. The left panel includes the AGN
sample (marked with $\times$ and $+$) of Sulentic et al. (2002). The
position of MWC 560 and CH Cyg is marked with open triangles. As
could be expected, they are located outside of the AGN population,
but close to  the NLSy1 galaxies (marked  with $+$), which are
supposed to have systematically lower BH masses. The diagram of
Fig.2 is believed to play for AGN role similar to the one of
Hertzsprung-Russell diagram for stars (Sulentic 2001). The physical
drivers can be L/M ratio convolved with orientation (Marziani et al.
2001), or  L/L$_{Edd}$ ($\propto$ L/M)  and  BH mass (Boroson 2002).
Here we want to use the ``nano-quasars" to understand better this
correlation space. The "nano-quasars" give us the unique possibility
to consider the effect  of L/M along with that of varying the  mass
seven-eight orders of magnitude.

The reverberation mapping  studies of the AGN (Kaspi et al 2001)
give the  following dependence of FWHM(\hbbc) on mass and L/M:
\begin{equation}
 FWHM(H\beta) = 4350 \left( \frac{L}{M} \right)_{su} ^{-0.35}
 \left( \frac{M}{M_\odot} \right)^{0.15}
 \;  \;  km \; s^{-1},
\end{equation}
where $(L/M)_{su}$ is the luminosity-to-mass ratio in solar units,
with the solar value (L/M)$_\odot$=1.92~ergs~s$^{-1}$g$^{-1}$.

The distance of the BLR from the central continuum source is found to
depend on bolometric luminosity L:
\begin{equation}
r = 9.36 \times 10^8 \left( \frac{L}{L_\odot} \right)^{0.7} \; \;  cm  \; .
\end{equation}

The ionization parameter can be defined as:
\begin{equation}
  U = \frac{Q(H)}{4 \pi c r^2 n_e }  \; ,
\end{equation}
where $Q(H)=fL/h\nu$, is the number of the hydrogen-ionizing
photons, $r$ is the distance of the broad line region from the
central continuum source, $f$\ is the fraction of ionizing photons,
$\nu$ the average frequency of the ionizing photons, and $n_e$ the
electron density. The appearance of the same lines in quasars and in
symbiotic stars means that the density in the line emitting region
should be similar. A relation connecting $n_e$\ and the (L/M)-ratio
is (Marziani et al. 2001):
\begin{equation}
n_{e} = 5.248 \times 10^7 \left( \frac{L}{M} \right)_{su} ^{\frac{2}{3}} \; cm^{-3} \;  .
\end{equation}
Combining the above  equations we obtain:
\begin{equation}
U  = 3.3 \times 10^6 \: f \: \left( \frac{ 10^{16} Hz }{\nu} \right)
\left( \frac{L}{M} \right)_{su} ^{-1.07}
M_{su} ^{-0.4}
\end{equation}
A typical AGN continuum (Laor et al. 1997) yields
$\nu\approx$1.22$\times$10$^{16}$\ Hz and $f \approx$0.39.
We can then calculate \rfe\ following  Marziani et al. (2001).
On the right panel of \ref{Fe-Hb} are shown  theoretical lines
covering the range of masses expected for AGN,
$\log$ (M/M$_\odot$) $\approx$ 6 -- 9.5.

In symbiotic and symbiotic-like stars the ionizing photons are coming
from accreting white dwarf, which effective temperature could
be in the range from 6000~K up to $\ga$200~000~K
(M\"urset \& Nussbaumer 1994). In CH~Cyg and MWC~560 the
UV continuum shape indicates a region of extremely large
column density (``cocoon")
that obscures the inner disk and the hot white dwarf surface
(Michalitsianos et al. 1993). The UV continuum fitting
of CH~Cyg during the active phase indicate temperature
T$_{eff}$=8500 -- 15000~K and temperature is the lowest
at the maximum of brightness(Mikolajewska et al. 1988).
Because the jet ejection is at the maximum, we will adopt
T$_{eff}$=8500~K corresponding to
$f=1 \times 10^{-5}$ and $\nu=3.48 \times 10^{15}\:$ Hz.
For MWC~560 there are also no signs for temperature hotter
than 15000$K$ (Shore et al. 1994), and even the
column density of the absorbing material is higher than in
CH~Cyg (Michalitsianos et al. 1993).

The resulting  FWHM(H$\beta$) vs. R$_{FeII}$ is plotted in Fig. 2.
On the left panel the lines are for masses M$_{BH}=5\times10^8$~\Ms,
M$_{BH}=10^7$~\Ms, and white dwarf mass M$_{WD}=1.4$~\Ms. We adopted
$f$ and $\nu$ following the above considerations. The $L/M$\ ratio
was running in the limits log($L/M)_{su}=$2.5 -- 4.4 for
M$_{BH}=1\times10^9$\Ms, log($L/M)_{su}=$2.5 -- 4.9 for
M$_{BH}=10^7$\Ms, and log($L/M)_{su}=$3.0-3.9 for M$_{WD}=1.4$\Ms.

The total luminosity of the white dwarf of CH~Cyg during the time
of  jet ejection in 1984 is L$\la$1600~L$_{\odot}$ (Mikolajewska et
al 1988) and that  of  MWC~560 is L$\approx $1000~L$_{\odot}$ (Schmid
et al. 2001). Assuming a typical  white dwarf mass  in symbiotic
stars of M$_{WD}=1.0-1.4$\Ms (or total mass of the binary system
about 3-5\Ms),  we obtain $(L/M)_{su}\approx$10$^3$ \ in agreement
with parameters used to plot the lowest line in Fig. 2 (left panel).

If we use higher effective temperature for the ionizing continuum
(i.e. T$_{eff}=15\:000^0 K$ corresponding to $f=6.61 \times 10^{-3}$
and $\nu=3.66 \times 10^{15}\:$ Hz) to have reasonable results from
Eq.5 we need to go to stronger dependence like to $U\propto
(L/M)^{-(1+x)} M^{-1}$ with $x$ up to $x\approx$0.67 (more details
are given in Marziani et al. 2001). This points out that the number
of ionizing photons and the shape of the UV and X-ray continuum are
important in the E1 correlations. It is worth noting that the soft
X-ray photon index is the third axis in E1 diagrams of Sulentic et
al. (2000, 2001). The line widths, FWHM(H$\beta$)$=$100~\kms \ for
MWC~560 and FWHM$\approx$200~\kms \ for CH~Cyg, are a little bit
smaller than expected from  Eq. 1, which predicts values around
300-400 \kms. It is also  possible that the luminosity of CH~Cyg and
MWC~560 was lower (i.e. 100-500 L$_\odot$) at the time the
spectra were obtained, if the distances used in the calculations are
too large. We regard the agreement acceptable, considering the huge
difference in masses and the limited range of luminosity and masses
from which the relationships of Kaspi et al. (2000) and Marziani et
al. (2001) were derived. The white dwarfs have lower efficiency of
accretion ($\eta \approx 10^{-4}$) than expected for typical AGN
($\eta \approx 10^{-1}-10^{-2}$). The used equations give good
results without considering the effect of efficiency. This is in
agreement with the interpretation of the E1 as mainly driven by L/M
ratio. Efficiency, along with several other parameters, may have
some (minor) influence.

Following the above equations and Fig. 2,  a change in mass by a
factor $\sim$10$^7$ times changes  FWHM(H$\beta$) by a factor 10 --
50. A change in mass by a factor 50 (from $10^7$ to 5$\times 10^8$
\Ms, in the range expected for BH masses of low redshift quasars)
leads to a FWHM(\hb) change by a factor 2-3. 
In a sample of quasars with a limited
BH mass spread, mass change may lead to second order effect which
may even be smaller than the the effect of orientation (Marziani et
al. 2001).

\section{Jets \label{jets}}

Jets are detected in systems quite different from the ones harboring
black holes (for a review see Livio, 2001): young stellar objects,
planetary nebulae, supersoft X-ray sources. The jets of MWC~560 and
CH~Cyg, as well as  other non-relativistic jets, are visible in
atomic spectral lines. Among the relativistic jets only in SS~433 are
atomic lines visible (Fender 2001). The jet velocities observed in
the accreting white dwarfs we call ``nanoquasars" are  $\approx$1000
\kms\  in CH~Cyg (Taylor et al. 1986) and 1000$-$6000 \kms\ in
MWC~560. These velocities are much slower than the jet velocities in
microquasars: 0.26c in SS~433 (Margon 1984), 0.5c in Cyg~X-3
(Mart\'\i\ et al., 2001), and 0.9c in GRS~1915+105 (Mirabel \&
Rodriguez 1999). However, they are consistent with an overall
picture in which the jet velocity is of the order of the escape
velocity (Livio 2001). A rough estimation is
v$_{esc}$(WD)$\approx$$0.02c$.

The compact object luminosity of MWC~560 and CH~Cyg is
considerably less than the Eddington limit $L \leq 0.05 L_{Edd}$.
At such accretion luminosity the
most probable jet energy source  is extraction of rotational energy
from the compact object. In the cases of ``nano-quasars" the
extraction is probably going on via the propeller action of a
magnetic white dwarf (Mikolajewski et al. 1996). The most probable
source of jet formation in quasars is the extraction of energy and
angular momentum via the Blandford \& Znajek (1977) mechanism. In
this sense the jets in the ``nano-quasars"  represent probably a low
energy (non-relativistic) analog of the jets of quasars and
microquasars, having a similar energy source -- the extraction of
rotational energy from the central compact object.

\section{Conclusions}

We showed a clear similarity between the emission line spectra  of
the accreting white dwarfs CH~Cyg and MWC~560 and low redshift
quasars. Their jets may be have a similar energy source - the
rotational energy of the central object. The discussed similarities
have been used in this paper to aid the interpretation of  the E1
diagrams for AGN. Their position in the diagram confirms the
interpretation that the physical driver of the Boroson \& Green's E1
is primarily  the L/M ratio, and that the mass of the accreting
object also plays a role. We could call the accreting white dwarfs
with jets and with AGN-like spectra ``nano-quasars", by analogy with
the quasar and microquasar denomination (also, $\nu \alpha \nu  o
\varsigma $ in ancient Greek means dwarf).

In the future, the spectral similarity discussed here could be used
\ to better understand  the conditions and the structure of the BLR
of quasars, and  to identify numerous emission lines in the spectrum
of I Zw 1  (and by extension, of the wide majority of AGN that show a
\feii\ spectrum almost identical to the one of I Zw 1). The FWHM of
emission lines in AGN is $\geq$1000, which makes most \feii\ lines
blended together. In accreting white dwarfs the FWHM is a factor of
ten less, making possible to identify weak lines. The diversity of
symbiotic binaries properties could be used to test and to
``calibrate" AGN correlations, using objects with well known mass.
It will be very useful to detect a stellar mass black hole
 (i.e., a galactic micro-quasar) with similar emission lines.  Discovery of an
 interacting binary in which a black hole accretes from the wind of a  red
 giant will be extremely interesting, although very difficult  to find
 from an evolutionary  point of view.

Ultimately, the nano-quasars could be  an important chain link in our
understanding of  all  accreting sources. They could help us to
create an unified picture of all accreting objects from cataclysmic
variables and stellar-mass black holes up to the most powerful
quasars.

\begin{acknowledgements}
We are grateful of the referee for useful comments.
This work is supported by Italian Ministry of University
and Scientific and technological Research (MURST) through
Cofin grant 00-02-004.
\end{acknowledgements}

\end{document}